\documentclass{Interspeech2024}

\interspeechcameraready 

\usepackage{hyperref}
\usepackage{verbatim}
\usepackage{times}
\usepackage{latexsym}
\usepackage{soul}
\usepackage[T1]{fontenc}
\usepackage{verbatim}
\usepackage{booktabs}
\usepackage{multirow}
\usepackage{graphicx}
\usepackage{amsmath}
\usepackage{subcaption}
\usepackage[most]{tcolorbox}
\title{The Reasonable Effectiveness of Speaker Embeddings for Violence Detection
}

\name[affiliation={1}]{Sarthak}{Jain*}
\name[affiliation={2}]{Orchid}{Chetia Phukan*}
\name[affiliation={2}]{Arun}{Balaji Buduru}
\name[affiliation={2,3}]{Rajesh}{Sharma}

\address{
  $^1$GGSIPU, New Delhi, India,
  $^2$IIIT-Delhi, India,
  $^3$University of Tartu, Estonia\\ *equal contribution}
\email{sarthakjainssjj@gmail.com, orchidp@iiitd.ac.in}


\begin{document}

\maketitle

\begin{abstract}
In this paper, we focus on audio violence detection (AVD). AVD is necessary for several reasons,
especially in the context of maintaining safety, preventing harm, and ensuring security in various environments. This calls for accurate AVD systems. Like many related applications in audio processing, the most common approach for improving the performance, would be by leveraging self-supervised (SSL) pre-trained models (PTMs). However, as these SSL models are very large models with million of parameters and this can hinder real-world deployment especially in compute-constraint environment. To resolve this, we propose the usage of speaker recognition models which are much smaller compared to the SSL models. Experimentation with speaker recognition model embeddings with SVM \& Random Forest as classifiers, we show that speaker recognition model embeddings perform the best in comparison to state-of-the-art (SOTA) SSL models and achieve SOTA results. 

\begin{comment}

 Like Many other applications in speech processing, the best/commonly approached for improving the performance, speech SSL embeddings are leveraged, however, these SSL PTMs are heavier with million of parameters, this hinders real-world deployment, in this work, we give an alternative to the use of SSL embeddings as input features by leveraging x-vector features which are speaker recognition features, x-vector being very light. for real world detecting violence with higher accuracy is needed and here comes the need of pre-trained features, however, the SOTA SSL models are too heavy to be deployed to real time systems and high inference time,,which makes them unfit for this..however, the spectral features like MFCC doesn't provide much accuracy, however being lightweight. SO in this work, we propose to use, XVector trained for speaker recognition but achieve SOTA performance in comparison to SSL counterparts only around with 7 M params.
 \end{comment}
\end{abstract}

\vspace{-0.2cm}

\section{Introduction}
%\noindent\textcolor{red}{O to S: 1st para: Start with writing negative effects of violence} \newline
%Violence can have numerous negative effects on individuals, communities, and society as a whole. Some of the key negative consequences associated with violence are physical harm, psychological impact, social disruption, economic consequences, and so on. 
Violence not only inflicts immediate harm but also leaves lasting scars on the fabric of society. The ripple effects extend beyond individual victims, impacting families, neighborhoods, and entire communities. As such, ensuring public safety and security becomes paramount, requiring proactive measures to address violence and foster environments where all members can thrive without fear of harm or injustice. 
In response to the negative effects associated with violence, previous research has made substantial leap towards violence detection through use of various modalities. %Researchers have experimented with visual \cite{rendon2023crimenet} as well as audio input \cite{he2024fcc} for violence detection. 
In this work, we focus on audio violence detection (AVD). AVD has advantages in conditions such as in areas with low visibility or at night for detecting violence, in comparison to visual or audio-visual violence where access to a visual monitoring device will be required. Researchers have explored AVD with methods ranging from classical machine learning to deep learning \cite{yildiz2023novel, bakhshi2023violence}. %Yildiz et al. \cite{yildiz2023novel, bakhshi2023violence} explored tunable Q wavelet transform with kNN and SVM as classifiers for AVD. Further, Bakshi et al. \cite{bakhshi2023violence} explored different vision pre-trained models with mel-spectrograms. Zhu et al. \cite{zhu2023computationally} experimented with shallow networks that consists of single hidden layer for lightweight AVD with various features such as MFCCs, $\Delta$MFCCs, zero-crossing rate and so on. 
However, despite all these advancements AVD still lacks in accuracy and considering the importance of AVD, systems with higher accuracy are need of the hour. \par

Following related speech and audio processing applications, the most common and beneficial way to boost up a system performance would be using embeddings from pre-trained models (PTMs) especially self-supervised (SSL) PTMs \cite{yang21c_interspeech}. However, these state-of-the-art (SOTA) SSL PTMs are very large with millions of parameters that may hinder real-world deployment of AVD systems especiallly in scenarios with less compute power. In response, in this work, we propose to use embeddings extracted from speaker recognition PTMs, models pre-trained primarily for speaker recognition for AVD which is much smaller in size than the SSL PTMs. With comprehensive comparative analysis, we show that x-vector (speaker recognition) embeddings attains the topmost performance. This topmost behavior of x-vector embeddings can be attributed to its effectiveness in capturing intensity, intonation, etc more effectively than SSL PTMs for AVD.

% \begin{figure*}[htbp]
%     \centering
%     \includegraphics[width=\textwidth]{12.pdf}
%     \caption{Audio Violence Detection Application; Application architecture pipeline is presented in Figure \ref{fig:my_label} (a) with the flow of information when the input audio is provided to the final inference received by the user through the user interface; Figure 1(c) shows the confusion matrix of the best model RF with x-vector embeddings}
%     \label{fig:my_label}
% \end{figure*}

\begin{figure*}[htbp]
    \centering
    
    \includegraphics[width=0.92\textwidth]{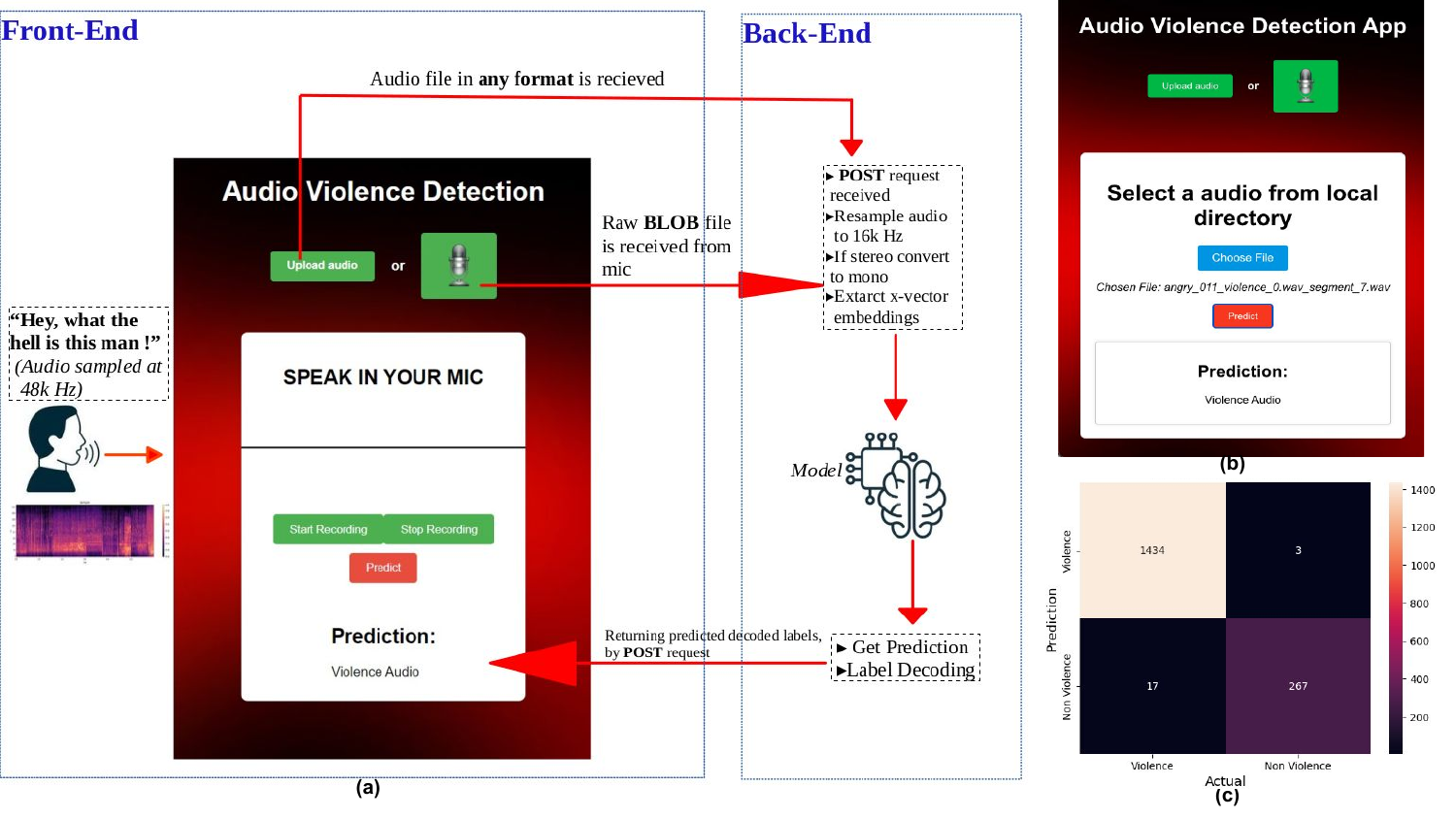}
    \caption{Audio Violence Detection Application; Application architecture pipeline is presented in Figure \ref{fig:my_label} (a) with the flow of information when the input audio is provided to the final inference received by the user through the user interface; Figure 1(c) shows the confusion matrix of the best model RF with x-vector embeddings}
    \label{fig:my_label}
\end{figure*}

% \begin{figure*}[htbp]
%     \centering
%     \makebox[\textwidth][c]{\includegraphics[width=1.0\textwidth]{12.pdf}}
%     \caption{Your Caption here}
%     \label{fig:my_label}
% \end{figure*}

\vspace{-0.3cm}

\section{Application}
In this section, we discuss various parts of the proposed AVD application. Firstly, we discuss the various PTMs whose embeddings are used in our study and the classifiers. Followed by the database considered and the results of the experiments. Lastly, we discuss the workings of the application and how a user can interact with the application.  

\noindent\textbf{Pre-Trained Models}: We use SOTA SSL PTMs WavLM \cite{9814838} and Unispeech-SAT \cite{9747077} because of its performance in various applications in SUPERB. For speaker recognition PTMs, we choose x-vector \cite{8461375} and ECAPA \cite{desplanques20_interspeech} due to their SOTA performance in speaker recognition. The number of parameters of the PTMs are 94.7M, 94.68M, 7M, 20M for wavLM, Unispeech-SAT, x-vector, ECAPA respectively. We extract embeddings of 768 from the last hidden state of SSL PTMs through average pooling and for x-vector, we extract embeddings of dimension of 512. \par 
\noindent\textbf{Classifiers}: We experiment with Random Forest (RF) and Support Vector Machine (SVM) classifier with default parameters. We train the models with 5-fold where four folds will be used for training and fifth fold for testing. \par

\noindent\textbf{Database and Pre-Processing}: We use AVD database\footnote{\url{https://www.kaggle.com/datasets/fangfangz/audio-based-violence-detection-dataset}} used by Zhu et al. \cite{zhu2023computationally}. %The database consists of audio files from many YouTube videos and represents violent incidents mostly captured using mobile phones.
We divided each of the violence and non violence audio files into sub audio files of 2.5 sec following \cite{zhu2023computationally} and results in dataset with non-violence audios of 7241 and violent audios of 1374. We resample the audios to 16kHz for passing as input to the PTMs. \par

% \begin{table}[ht]
% \caption{Performance Scores; Scores are average of 5 folds; Scores are presented in \%; F1 is macro-average F1-Score; MFCCs are used as baseline}

% \tiny
% \centering
% \begin{tabular}{l|c|c}
% \toprule
% \textbf{\textcolor{red}{PTM}} & \textbf{\textcolor{red}{Accuracy}} & \textbf{\textcolor{red}{F1}} \\
% \midrule

% %---------------------------------------------------%

% \multicolumn{3}{c}{\textbf{\textcolor{blue}{Random Forest}}} \\
% \midrule
% x-vector & \textbf{99.14} & \textbf{98.40}\\

% ECAPA &  94.01 & 87.42\\

% MFCC & 72.46 & 49.01 \\

% Unispeech-SAT & 94.07  &  87.99   \\

% WaveLM &   91.05 &  80.61 \\

% \midrule

% %---------------------------------------------%

% \multicolumn{3}{c}{\textbf{\textcolor{blue}{Support Vector Machine}}} \\
% \midrule
% x-vector & 99.13 &  98.38 \\

% ECAPA &  98.02 & 98.02\\

% MFCC & 71.93 & 48.71 \\

% Unispeech-SAT &  97.73  & 95.87  \\
% WaveLM & 95.53 & 91.40  \\

% \bottomrule
% \end{tabular}
% \label{tableacc}
% \end{table}

\begin{table}[ht]
\caption{Performance Scores; Scores are average of 5 folds; Scores are presented in \%; F1 is macro-average F1-Score; MFCCs are used as baseline}

% Adjust font size to 22.5pt
\fontsize{6.5}{4}\selectfont

\centering
\begin{tabular}{l|c|c}
\toprule
\textbf{\textcolor{red}{PTM}} & \textbf{\textcolor{red}{Accuracy}} & \textbf{\textcolor{red}{F1}} \\
\midrule

%---------------------------------------------------%

\multicolumn{3}{c}{\textbf{\textcolor{blue}{Random Forest}}} \\
\midrule
x-vector & \textbf{99.14} & \textbf{98.40}\\

ECAPA &  94.01 & 87.42\\

MFCC & 72.46 & 49.01 \\

Unispeech-SAT & 94.07  &  87.99   \\

WaveLM &   91.05 &  80.61 \\

\midrule

%---------------------------------------------%

\multicolumn{3}{c}{\textbf{\textcolor{blue}{Support Vector Machine}}} \\
\midrule
x-vector & 99.13 &  98.38 \\

ECAPA &  98.02 & 98.02\\

MFCC & 71.93 & 48.71 \\

Unispeech-SAT &  97.73  & 95.87  \\
WaveLM & 95.53 & 91.40  \\

\bottomrule
\end{tabular}
\label{tableacc}
\end{table}

\noindent\textbf{Experimental Results}: The evaluation results of the various models are presented in Table \ref{tableacc}. Models trained with speaker recognition models (x-vector, ECAPA) embeddings performed the best. This can be attributed to their supremacy in capturing speech characteristics such as intensity, intonation, etc more effectively. We use the best model Random Forest trained with x-vector embeddings as backend model in the application. 
% \begin{figure}[t!]
%     \centering
%     \subfloat[x-vector]{{\includegraphics[width=0.52\linewidth, height = 0.8\linewidth ]{11.png}}\label{aaa}}
%     \subfloat[ECAPA]
%     {{\includegraphics[width=0.52\linewidth, height = 0.8\linewidth ]{1.png}}\label{apa}}\\
%    \caption{User Interface; Figure \ref{aaa}, Figure \ref{apa} represents the two-modes: user can record audio or upload a audio}
% \label{fig:tsne}
% \end{figure}

% \begin{figure}
%     \centering
%     \includegraphics[width=1\linewidth]{10.png}
%     \caption{Architecture Pipeline}
%     \label{fig:label}
% \end{figure}

\noindent\textbf{User Interface: Working and Backend Details}: The user interface of the application is presented in Figure \ref{fig:my_label}. %Also, Figure \ref{aaa}, Figure \ref{apa} represents the two settings a user can interact with the application. 
In mode 1 (Figure \ref{fig:my_label} (a)), user can use \textit{Start Recording} and \textit{Stop Recording} button to record real-world audio for AVD and in mode 2 (Figure \ref{fig:my_label} (b)) user can upload a audio with \textit{Choose File} button for detecting violence in the uploaded audio. With \textit{Predict} button, we can get the predictions out of the application. We have used React.Js and Flask for creating the Front-End and the Back-End for model inference respectively. The average inference time for a 1 min audio file is around 1 sec, thus making it deployable to real-world use.

\vspace{-0.7cm}
\section{Conclusion}

In this demonstration, we've created an application that harnesses speaker recognition model embeddings to enhance AVD. The application streamlines the process of detecting violent activities across surveillance feeds without access to visual-monitoring device . Overall, the application offers law enforcement agencies a powerful tool for more efficient and effective monitoring, ultimately enhancing public safety.
\bibliographystyle{IEEEtran}
\bibliography{main.bib}

\begin{thebibliography}{1}
\providecommand{\url}[1]{#1}
\csname url@samestyle\endcsname
\providecommand{\newblock}{\relax}
\providecommand{\bibinfo}[2]{#2}
\providecommand{\BIBentrySTDinterwordspacing}{\spaceskip=0pt\relax}
\providecommand{\BIBentryALTinterwordstretchfactor}{4}
\providecommand{\BIBentryALTinterwordspacing}{\spaceskip=\fontdimen2\font plus
\BIBentryALTinterwordstretchfactor\fontdimen3\font minus \fontdimen4\font\relax}
\providecommand{\BIBforeignlanguage}[2]{{%
\expandafter\ifx\csname l@#1\endcsname\relax
\typeout{** WARNING: IEEEtran.bst: No hyphenation pattern has been}%
\typeout{** loaded for the language `#1'. Using the pattern for}%
\typeout{** the default language instead.}%
\else
\language=\csname l@#1\endcsname
\fi
#2}}
\providecommand{\BIBdecl}{\relax}
\BIBdecl

\bibitem{yildiz2023novel}
A.~M. Yildiz, P.~D. Barua, S.~Dogan, M.~Baygin, T.~Tuncer, C.~P. Ooi, H.~Fujita, and U.~R. Acharya, ``A novel tree pattern-based violence detection model using audio signals,'' \emph{Expert systems with applications}, vol. 224, p. 120031, 2023.

\bibitem{bakhshi2023violence}
A.~Bakhshi, J.~Garc{\'\i}a-G{\'o}mez, R.~Gil-Pita, and S.~Chalup, ``Violence detection in real-life audio signals using lightweight deep neural networks,'' \emph{Procedia Computer Science}, vol. 222, pp. 244--251, 2023.

\bibitem{yang21c_interspeech}
S.~wen Yang, P.-H. Chi, Y.-S. Chuang, C.-I.~J. Lai, K.~Lakhotia, Y.~Y. Lin, A.~T. Liu, J.~Shi, X.~Chang, G.-T. Lin, T.-H. Huang, W.-C. Tseng, K.~tik Lee, D.-R. Liu, Z.~Huang, S.~Dong, S.-W. Li, S.~Watanabe, A.~Mohamed, and H.~yi~Lee, ``{SUPERB: Speech Processing Universal PERformance Benchmark},'' in \emph{Proc. Interspeech 2021}, 2021, pp. 1194--1198.

\bibitem{9814838}
S.~Chen, C.~Wang, Z.~Chen, Y.~Wu, S.~Liu, Z.~Chen, J.~Li, N.~Kanda, T.~Yoshioka, X.~Xiao, J.~Wu, L.~Zhou, S.~Ren, Y.~Qian, Y.~Qian, J.~Wu, M.~Zeng, X.~Yu, and F.~Wei, ``Wavlm: Large-scale self-supervised pre-training for full stack speech processing,'' \emph{IEEE Journal of Selected Topics in Signal Processing}, vol.~16, no.~6, pp. 1505--1518, 2022.

\bibitem{9747077}
S.~Chen, Y.~Wu, C.~Wang, Z.~Chen, Z.~Chen, S.~Liu, J.~Wu, Y.~Qian, F.~Wei, J.~Li, and X.~Yu, ``Unispeech-sat: Universal speech representation learning with speaker aware pre-training,'' in \emph{ICASSP 2022 - 2022 IEEE International Conference on Acoustics, Speech and Signal Processing (ICASSP)}, 2022, pp. 6152--6156.

\bibitem{8461375}
D.~Snyder, D.~Garcia-Romero, G.~Sell, D.~Povey, and S.~Khudanpur, ``X-vectors: Robust dnn embeddings for speaker recognition,'' in \emph{2018 IEEE International Conference on Acoustics, Speech and Signal Processing (ICASSP)}, 2018, pp. 5329--5333.

\bibitem{desplanques20_interspeech}
B.~Desplanques, J.~Thienpondt, and K.~Demuynck, ``{ECAPA-TDNN: Emphasized Channel Attention, Propagation and Aggregation in TDNN Based Speaker Verification},'' in \emph{Proc. Interspeech 2020}, 2020, pp. 3830--3834.

\bibitem{zhu2023computationally}
F.~Zhu-Zhou, D.~Tejera-Berengu{\'e}, R.~Gil-Pita, M.~Utrilla-Manso, and M.~Rosa-Zurera, ``Computationally constrained audio-based violence detection through transfer learning and data augmentation techniques,'' \emph{Applied Acoustics}, vol. 213, p. 109638, 2023.

\end{thebibliography}

\begin{comment}

\begin{figure}
    \centering
    \includegraphics[width=0.5\linewidth]{3.png}
    \caption{User speaking in mic}
    \label{fig:label}
\end{figure}

\begin{figure}
    \centering
    \includegraphics[width=0.8\linewidth]{1.png}
    \caption{User selecting an audio file}
    \label{fig:label}
\end{figure}

\end{comment}

\end{document}